\documentclass[english,12pt,a4paper]{article}
\pdfoutput=1

\usepackage{authblk}
\usepackage[text={17.0cm,23.7cm}]{geometry}
\usepackage[T1]{fontenc}
\usepackage[utf8]{inputenc}
\usepackage{babel}
\usepackage{hyperref}
\usepackage{cite}
\usepackage{graphicx}
\usepackage{amsmath}
\usepackage{amssymb}
\usepackage[dvipsnames]{xcolor}
\usepackage{yfonts}
\usepackage{ulem}
\usepackage{enumitem}
\usepackage{slashed}
\usepackage{textcomp}%for Zel'dovich name

\hypersetup{colorlinks,bookmarksopen,bookmarksnumbered,citecolor=blue,
	linkcolor=black,pdfstartview=FitH,urlcolor=blue}

\newcommand{\eV}[0]{\,\mathrm{eV}}
\newcommand{\keV}[0]{\,\mathrm{keV}}
\newcommand{\MeV}[0]{\,\mathrm{MeV}}
\newcommand{\GeV}[0]{\,\mathrm{GeV}}

\begin{document}
	
	\title{\vspace{-2cm}
		{\normalsize
			\flushright TUM-HEP 1245/19\\}
		\vspace{0.6cm}
		\textbf{Probing multicomponent FIMP scenarios with gamma-ray telescopes}\\[8mm]}
	
	\author[]{Johannes Herms}
	\author[]{Alejandro Ibarra}
	\affil{\normalsize\textit{Physik-Department T30d, Technische Universit\"at M\"unchen, James-Franck-Stra\ss{}e, 85748 Garching, Germany}}
	
	\date{}
	
	\maketitle\begin{abstract}
		We consider a scenario where the dark sector includes two Feebly Interacting Massive Particles (FIMPs), with couplings to the Standard Model particles that allow their production in the Early Universe via thermal freeze-in. These couplings generically lead to the decay of the heavier dark matter component into the lighter, possibly leading to observable signals of the otherwise elusive FIMPs. Concretely, we argue that the loop induced decay  $\psi_2\rightarrow\psi_1\gamma$ for fermionic FIMPs, or  $\phi_2\rightarrow\phi_1\gamma\gamma$ for scalar FIMPs, could have detectable rates for model parameters compatible with the observed dark matter abundance. 
	\end{abstract}
	
\renewcommand{\thefootnote}{\arabic{footnote}}
\newcommand{\bhline}[1]{\noalign{\hrule height #1}}
\newcommand{\bvline}[1]{\vrule width #1}

\setcounter{footnote}{0}

\setcounter{page}{1}

\section{Introduction}
For many years thermal freeze-out \cite{Zel_dovich_1966,Lee:1977ua} has been the most favored framework for particle dark matter production (for reviews, see {\it e.g.} \cite{Jungman:1995df,Bertone:2004pz,Bergstrom:2000pn,Feng:2010gw}). In this framework, dark matter particles are assumed to have interactions with the Standard Model particles which are strong enough to maintain both sectors in thermal equilibrium with each other, and weak enough to allow the dark matter population to leave thermal equilibrium sufficiently early. After this epoch, dark matter annihilations in the large-scale Universe are rare, leaving a relic population of dark matter particles which may account for observations. At small scales, however, there are regions with overdensities of dark matter particles where annihilations may have a sizable rate at the present epoch,
producing a potentially detectable flux of photons and other stable Standard Model particles.  Notably, the values of the coupling strengths required to reproduce the observed dark matter abundance predict fluxes for these cosmic ray species which could be large enough to be discerned from the astrophysical backgrounds, thus providing a test of the freeze-out mechanism. 

The absence of annihilation signals from the galactic center or from dwarf galaxies, while not excluding the freeze-out mechanism, has triggered interest in alternative dark matter production mechanisms. One of them is the so-called ``freeze-in'' mechanism \cite{Hall:2009bx}, which assumes dark matter interactions with the Standard Model which are too feeble to bring the dark matter into thermal equilibrium with our visible sector. These interactions, on the other hand, allow dark matter production through decays or collisions of Standard Model particles, which can lead to the observed dark matter abundance for appropriate parameters. It should be borne in mind that non-thermal production mechanisms occurring at very early times, such as inflaton decay, could also contribute to the total dark matter abundance, in contrast to the freeze-out case, where thermalization  erases any memory of the earliest stages of the cosmological history.

The feeble couplings to the Standard Model could also explain the longevity of the dark matter without invoking ad-hoc symmetries. The rare dark matter decays could lead in this framework to tests of the freeze-in mechanism. Current X-ray bounds require for this minimal scenario a dark matter mass lighter than a few tens of keV~\cite{Heeba:2018wtf}. For larger masses it is necessary to introduce an additional symmetry in order to suppress, partially or completely, the decays. In this case, testing the freeze-in mechanism becomes extremely challenging (see however \cite{Hambye:2018dpi}).  

In view of the complexity of our visible sector, it is conceivable that the dark sector could also be complex and that more than one FIMP exists in the particle spectrum, and that more than one FIMP is present in our Universe today produced via thermal freeze-in. In this scenario, the heavier FIMP could decay into the lighter one plus Standard Model particles, thus leading to new possibilities to test the freeze-in mechanism.  

In this work we consider scenarios with multicomponent scalar or fermionic FIMP dark matter, which couple to the same Standard Model fermion and to the same mediator through a Yukawa coupling, and in the case of the scalar FIMPs, also via the Higgs-portal. Scenarios along these lines of single component FIMP dark matter have been considered, {\it e.g.} in \cite{Garny:2018ali,Junius:2019dci,Belanger:2018mqt,Yaguna:2011qn,Molinaro:2014lfa}.
These couplings lead to FIMP production via freeze-in and allow the decay of the heavier FIMP. We consider in particular the one-loop induced decays $\psi_2\rightarrow\psi_1\gamma$ for fermionic dark matter and $\phi_2\rightarrow\phi_1\gamma\gamma$ for scalar dark matter, which generate a contribution to the photon flux with a very distinctive energy spectrum.
Such gamma ray spectral features can be easily separated from the featureless astrophysical background, making them a golden channel for positive dark matter identification (for reviews, see ~\cite{Bringmann:2012ez,Ibarra:2013cra}).
We will then investigate whether the decay rates expected from freeze-in production can be at the reach of current gamma-ray instruments.

This work is organized as follows:
In Section~\ref{sec:freezein}, we recapitulate some analytical results on freeze-in dark matter production. In Section~\ref{sec:fermionFIMP} we analyze the gamma-ray signals from multicomponent fermion FIMP dark matter, assuming FIMP production through the decay of a heavy exotic scalar, in Section \ref{sec:scalarFIMP-Higgs} from multicomponent scalar FIMPs, assuming FIMP production through the Higgs-portal and in Section \ref{sec:scalarFIMP-loop} from multicomponent scalar FIMPs, assuming FIMP production through the decay of a heavy exotic fermion. We present our conclusions in Section~\ref{sec:conclusion}.

\section{Freeze-in dark matter production}
\label{sec:freezein}

In the following, we recapitulate the main features of the freeze-in mechanism. We consider a FIMP dark matter candidate $\psi$ that couples to a Standard Model particle $X$ and to a heavy particle $\Sigma$ which we assume in thermal equilibrium with the plasma of Standard Model particles over the production process.
The time evolution of the dark matter number density $n_\psi$ is described by the Boltzmann equation \cite{Hall:2009bx}:
\begin{equation}
 \label{eq:BoltzmannGeneral}
 \frac{dn_\psi}{dt} + 3 H n_\psi = C_{1 \to 2} + C_{2 \to 2}\;,
\end{equation}
where $H$ is the Hubble expansion rate, while $C_{1 \to 2}$ and $C_{2 \to 2}$ are collision terms describing respectively $1 \to 2$ decay processes (such as $\Sigma \to \psi X$) and $2 \to 2$ scattering processes (such as $\Sigma X \to \psi X'$).
In the freeze-in scenario, the FIMP number density is much smaller than its equilibrium value during the whole thermal history. Hence, loss terms can be neglected and the collision terms can be cast as:
\begin{align}
 C_{1 \to 2} =& \sum_X \int \frac{d^3p_\Sigma}{(2\pi)^3} \frac{ g_\Sigma f_\Sigma \Gamma_{\Sigma\to X \psi}}{\gamma_\Sigma} \;,\\
 C_{2 \to 2} =& \sum_{a,b,X}
 \int 
 \frac{d^3p_a}{(2\pi)^3 2E_a}  \frac{d^3p_b}{(2\pi)^3 2E_b} 
 \frac{d^3p_\psi}{(2\pi)^3 2E_\psi} \frac{d^3p_X}{(2\pi)^3 2E_X} g_a f_a g_b f_b g_X (1 \pm f_X) g_\psi (1 \pm f_\psi) \nonumber \\
  &\qquad\qquad\qquad\qquad \left| \mathcal{M}_{a b \to \psi X} \right|^2 2\pi^4 \delta^4(p_a+p_b-p_\psi-p_X)\;,
\end{align}
where we have summed over all possible particles $a,b,X$ possibly involved in the process. Here, $g_X$ and $f_X$ are  the number of degrees of freedom and the phase space density distribution of the particle $X$, respectively, $\Gamma_{\Sigma\rightarrow X\psi}$ is the decay rate of the decay process $\Sigma\rightarrow X\psi$ and  $\gamma_\Sigma=E_\Sigma/m_\Sigma$ accounts for time dilation. For simplicity, we will assume that the phase space distributions for all particles except for the FIMP and the mediator follow a Maxwell-Boltzmann distribution; the effects of the Bose/Fermi enhancement/suppression factors for the FIMP relic abundance have been discussed in~\cite{Belanger:2018mqt} and can modify the results by ${\cal O}(1)$ factors.

When the decays $\Sigma \rightarrow \psi X$ are kinematically allowed, the $1\to 2$ collision term typically dominates over the $2\to 2$ term \cite{Hall:2009bx,Chu:2011be} (see however \cite{Junius:2019dci,Belanger:2018mqt}). In this case, the Boltzmann equation Eq.~(\ref{eq:BoltzmannGeneral}) can be written as \cite{Hall:2009bx}
\begin{equation}
 \label{eq:BoltzmannDecay}
 \frac{dY_\psi}{dx} = \sum_{X} \frac{g_\Sigma}{x H(T) s(T)}
 \Gamma_{\Sigma\to \psi X} 
 \int \frac{d^3p_\Sigma}{(2\pi)^3} \frac{m_\Sigma}{E_\Sigma} f_\Sigma(p_\Sigma,T) \;,
\end{equation}
where we have defined the yield $Y\equiv n/s$, with $s$ the entropy density, and the parameter $x\equiv m_\psi/T$, with $T$ the temperature of the thermal bath. For a radiation dominated Universe $H(T)=\sqrt{8 \pi^3/90} g^{1/2}_\mathrm{eff}(T) T^2/M_P$ and $s(T)=\frac{2 \pi^2}{45} g_\mathrm{eff}^s(T) T^3$, where $M_P$ is the Planck scale and  $g_\mathrm{eff}(T)$ and $g^{s}_\mathrm{eff}(T)$ are the effective number of degrees of freedom contributing respectively to the energy and the entropy density of the Universe at the temperature $T$~\cite{Gondolo:1990dk}. Assuming that the effective number of degrees of freedom in the Standard Model bath does not vary in the epoch where FIMP production is most efficient, $T_{\text{prod}}\sim {\cal O}(1)\times \mathop{\mathrm{max}}(m_\Sigma,m_\psi)$, one obtains a yield at the present epoch given by:
\begin{equation}
\label{eq:FreezeInAbundanceFactorised}
 Y_\psi^{\rm today} =
 \frac{g_\Sigma m_\Sigma^3}{H(m_\Sigma) s(m_\Sigma)}  \Gamma_{\Sigma\to \psi X} {\cal I_\pm}\;,
\end{equation}
where ${\cal I_\pm}$ is a dimensionless integral defined as
\begin{align}
{\cal I_\pm}=\int_0^\infty dx x^4 \int_1^\infty \frac{d\gamma}{2 \pi^2}  \frac{\sqrt{\gamma^2-1}}{e^{\gamma x} \pm 1} \;,
\end{align}
which takes numerical values ${\cal I}_+=0.248$ and ${\cal I}_-=0.232$ for $\Sigma$ a boson or a fermion. 
For a decaying boson, this yield results in:
\begin{equation}
\label{eq:freezeInAbundanceResult}
 \Omega_\psi = \left( \frac{\Gamma_{\Sigma\to \psi X}}{9.7 \times 10^{-25}\mathrm{GeV}}\right) \left( \frac{m_\psi}{\mathrm{GeV}}\right) g_\Sigma \left( \frac{m_\Sigma}{\mathrm{GeV}}\right)^{-2}
 \left(\frac{g_\mathrm{eff}(T_{\text{prod.}})}{106.75}\right)^{-3/2}\;.
\end{equation}
Comparing with the observed dark matter abundance, $\Omega_\mathrm{DM}h^2 = 0.120$~\cite{Aghanim:2018eyx} one obtains that the decay rate for the process $\Sigma\rightarrow \psi X$ giving a fraction of the total dark matter abundance $\Omega_\psi/\Omega_{\rm DM}$ is:
\begin{equation}
\label{eq:correctDecayRate}
 \Gamma_{\Sigma\to \psi X} = 1.2 \times 10^{-25} g_\Sigma^{-1} \frac{m_\Sigma^2}{m_\psi}
 \left(\frac{g_\mathrm{eff}(T_{\text{prod.}})}{106.75}\right)^{3/2}
 \frac{\Omega_{\psi}}{\Omega_\text{DM}}\;.
\end{equation}
This decay rate must be multiplied by a factor 1/2 if $X=\psi$, since in this case the decay $\Sigma\rightarrow \psi\psi$ produces two dark matter particles.

In some instances scattering processes can contribute significantly to FIMP production (see {\it e.g.} \cite{Hall:2009bx,Chu:2011be,Junius:2019dci}). For this case, the analytical treatment of the dark matter production becomes more complicated. Furthermore, and in contrast to freeze-out production, where the relic abundance is set at $T \lesssim m_\mathrm{DM}/20$, freeze-in production is most efficient around $T \sim \mathrm{max}(m_\Sigma, m_\psi)/\mathrm{few}$. Thus for $m_\psi\gtrsim 100\GeV$, electroweak symmetry restoration and the sizable thermal contributions to $m_\Sigma$ must be taken into account~\cite{Heeba:2018wtf}. 	In our work, we will analytically calculate the relic abundance using the formalism presented in this section. We have checked agreement with the numerical code micrOMEGAs\cite{Belanger:2018mqt}, employing FeynRules \cite{Alloul:2013bka} and CalcHEP \cite{Belyaev:2012qa} in the broken electroweak phase, extending our results into the scattering-dominated regime numerically where indicated.

\section{Multicomponent fermion FIMP DM from heavy scalar decay}
\label{sec:fermionFIMP}

We consider first a dark matter scenario consisting of two singlet Majorana fermions, $\psi_1$ and $\psi_2$, with masses $m_1$ and $m_2$, such that $m_2>m_1$.  We also impose a $\mathbb{Z}_2$ symmetry, unbroken in the electroweak vacuum, under which $\psi_1$ and $\psi_2$ are odd, while all Standard Model particles are even.  To couple them to the bath of Standard Model particles, we introduce a heavy scalar particle $\Sigma$, with mass $m_\Sigma>m_2, m_1$, also odd under the same $\mathbb{Z}_2$ symmetry, and with quantum numbers such that the Yukawa coupling $\bar X \psi_i \Sigma $ is allowed, with $X$ a Standard Model fermion. Being charged under the Standard Model gauge group, one generically expects $\Sigma$ to be in thermal equilibrium with the Standard Model bath. 

Let us assume for concreteness that $X$ is a right-handed lepton. The Lagrangian of the model then contains the following terms
\begin{align}
\mathcal{L}\supset
\left(\mathcal{D}_\mu \Sigma \right)^\dagger \left(\mathcal{D}^\mu \Sigma \right)
+ m_\Sigma \Sigma^\dagger \Sigma 
+ \lambda_{H\Sigma} |H|^2 |\Sigma|^2
+\Big( \frac{1}{2} \overline{\psi_i} i\slashed \partial \psi_i
- \frac{1}{2} m_i \overline{\psi_i^c} \psi_i
+ g_{ i} \bar l P_L \psi_i \Sigma + \mathrm{h.c.}\Big)\;.
\end{align}
We assume that the couplings $g_{i}$ are very small, such that $\psi_1$ and $\psi_2$ are FIMPs. In this framework, $\psi_1$ is the lightest $\mathbb{Z}_2$-odd particle and is absolutely stable. On the other hand, the heavier FIMP $\psi_2$ can decay into $\psi_1$ and Standard Model particles through a virtual $\Sigma$, either at tree level $\psi_2\rightarrow \psi_1 l^+ l^-$ or at the one loop level $\psi_2\rightarrow \psi_1 \gamma$ (and also $\psi_2\rightarrow \psi_1 Z, h$ when kinematically allowed). The decay rate is proportional to $\left|g_{1}\right|^2\left|g_{2}\right|^2$, therefore $\psi_2$ can be cosmologically long-lived. Finally, $\Sigma$ also decays, however with a cosmologically short lifetime, since the decay rate is only proportional to $\left|g_{i}\right|^2$. Yet, $\Sigma$ could be long lived enough to leave heavily ionizing charged tracks at the LHC detectors or to alter the abundances of primordial elements, if still present in significant amounts at the onset of Big Bang Nucleosynthesis. 

We will focus here on the prospects for observing gamma-ray signatures in the multicomponent fermionic FIMP dark matter scenario. The gamma-ray flux at Earth from the decay $\psi_2\rightarrow\psi_1\gamma$ can be calculated from the gamma-ray source term (see, {\it e.g.}, \cite{Ibarra:2013cra}),  which is defined as the rate of production of gamma-rays per unit energy inside the unit volume centered at the point $\vec r$:
\begin{align}
Q(E,\vec r)=\frac{\rho_{\psi_2}(\vec r)}{m_{\psi_2}} \Gamma_{\psi_2\rightarrow\psi_1\gamma}\frac{dN_\gamma}{dE}\;.
\label{eq:source-term}
\end{align}
Here $\rho_{\psi_2}(\vec r)$ is the mass density of the decaying dark matter component at the position $\vec r$, $dN_\gamma/dE$ is the energy spectrum of the photons produced in the decay, and $\Gamma_{\psi_2\rightarrow\psi_1\gamma}$ is the partial decay rate of the process $\psi_2\rightarrow\psi_1\gamma$, which reads~\cite{Garny:2010eg},
\begin{equation}
\label{eq:fermionDecayRateLowMassLimit}
\Gamma_{\psi_2\to\psi_1\gamma} = \frac{e^2 |g_{1} g_{2}|^2}{2^{15} \pi^5}\frac{m_{\psi_2}^5}{m_\Sigma^4} \left(1-\frac{m_1^2}{m_2^2} \right)^3 \left(1-\frac{m_1}{m_2} \right)^2.
\end{equation}

We will assume in what follows that the fraction of the dark matter mass density in the form of the unstable component $\psi_2$ is the same at all positions, and in particular the same to the value in the Universe at large scale: $\rho_{\psi_2}(\vec r)=\rho_{\rm DM} (\vec r)
\Omega_{\phi_2}/\Omega_{\rm DM}$.  The gamma-ray flux at Earth from a given direction can then be calculated using Standard tools (see, {\it e.g.} \cite{Ibarra:2013cra}), and receives contributions from decays of dark matter particles in our galaxy and of dark matter particles in the large-scale Universe.

In the freeze-in scenario, the population of $\psi_1$ and $\psi_2$ in the Universe is generated in the decays $\Sigma\rightarrow \psi_i l$,  leading to a relic abundance which is given by  Eq.~(\ref{eq:freezeInAbundanceResult}), with 
\begin{align}
\Gamma_{\Sigma\to \psi_i \bar l} &= \frac{1}{16 \pi m_\Sigma} |g_{i}|^2 \left(m_\Sigma^2 - (m_l^2+m_i^2)\right) \sqrt{1-\frac{2(m_i^2+m_l^2)}{m_\Sigma^2}+\frac{2(m_i^2-m_l^2)^2}{m_\Sigma^4}} \\
&\simeq \frac{m_\Sigma |g_{i}|^2}{16 \pi}.
\end{align}
The  gamma-ray flux at Earth then depends on the couplings constants $g_{i}$ through the decay rate and through the abundance of the heavier dark matter component, namely on $|g_{1}|^2|g_{2}|^4$. On the other hand, the coupling constants are constrained by the requirement that the FIMP density does not exceed the measured dark matter density. Therefore, one expects an upper limit on the gamma-ray flux from avoiding dark matter overabundance. 

Using the results of Section \ref{sec:freezein}, one finds that the requirement $\Omega_{\psi_1}+\Omega_{\psi_2}\leq \Omega_{\rm DM}$ translates into:
\begin{align}
|g_{1}|^2 \frac{m_1}{m_\Sigma} + |g_{2}|^2 \frac{m_2}{m_\Sigma} \leq 2.9 \times 10^{-24}\;.
\label{eq:fermion_relic_abundance}
\end{align}
This limit on the couplings is conservative, as the relic abundances of $\psi_{1,2}$ also receive contributions from freeze-in through scattering, and from super-WIMP production through out-of-equilibrium decay of frozen-out $\Sigma$ particles. However, for the range of adopted parameters $m_\Sigma \gg m_{2,1}$, one finds that the scattering contribution is subdominant \cite{Junius:2019dci}, and we find the same for the super-WIMP contribution.

Using Eq.~(\ref{eq:fermionDecayRateLowMassLimit}) and 
Eq.~(\ref{eq:correctDecayRate})
we obtain the following upper limit on the decay width for $\psi_2\rightarrow\psi_1\gamma$ from freeze-in production:
\begin{equation}
\Gamma_{\psi_2 \to \psi_1 \gamma}
\lesssim
	\left( 8 \times 10^{30} \;\mathrm{s}  \right)^{-1}
    \left(1-\frac{m_1^2}{m_2^2} \right)^3
    \left(1-\frac{m_1}{m_2} \right)^2
    \left(\frac{m_{\psi_1}}{\GeV}\right)^{-1}
	\left(\frac{m_{\psi_2}}{\GeV}\right)^4
	\left(\frac{m_\Sigma}{\GeV}\right)^{-2}
	\frac{\Omega_{\mathrm{\psi_2}}}{\Omega_\mathrm{DM}}
	\frac{\Omega_{\mathrm{\psi_1}}}{\Omega_\mathrm{DM}}\;.
\end{equation}
Since the source term is $Q(E,\vec r)\propto \Gamma_{\psi_2\rightarrow\psi_1\gamma}\Omega_{\psi_2}$, this upper limit on the rate translates into an upper limit on the source term, and correspondingly on the gamma-ray flux at Earth. The upper limit is saturated for $\Omega_{\psi_2}=2\Omega_{\psi_1}$, which corresponds to the values of the couplings
\begin{align}
\label{eq:optimalCouplings}
g_{2} &= 1.4 \times 10^{-12} \sqrt{m_\Sigma/m_2}\;, \nonumber\\
g_{1} &= 9.9 \times 10^{-13} \sqrt{m_\Sigma/m_1}.
\end{align}

Let us remark that in our FIMP scenario $\psi_1$ and $\psi_2$ are assumed to be out of thermal equilibrium over the whole cosmological history, which implies an upper limit on the coupling constant.   Using $Y_i^{\mathrm{fi}} < Y_i^\mathrm{eq} \sim \frac{T^3 g_i /\pi^2}{g_\mathrm{eff}(T) T^3 2 \pi^2 / 45}$, we obtain:
\begin{equation}
	\label{eq:thermalisationLimitg}
	|g_{i}| \lesssim 5\times 10^{-9} \left( \frac{m_\Sigma}{\GeV}\right)^{1/2}\;,
\end{equation}
which using Eq.~(\ref{eq:optimalCouplings}) translates into the lower limits on the FIMP masses $m_1 \gtrsim 40 \eV$,  $m_2 \gtrsim 80 \eV$.

The FIMP mass is also bounded by dark matter warmness constraints.
	In the single component freeze-in scenario, small-scale structure constraints \cite{Irsic:2017ixq} on the free-streaming of dark matter can be cast into the bound on the FIMP mass, $m_\mathrm{FIMP} \geq 15.6 \keV \times \left(106.75 / g_{\mathrm{eff}s}(T_{\text{prod.}})\right)^{1/3}$ \cite{Kamada:2019kpe}.
	We will always assume $\psi_2$ to be heavier than this value, to 
	ensure the existence of at least one cold dark matter component. On the other hand, observations do not preclude a fraction of the dark matter to be warm or hot, therefore we will only require for $\psi_1$ to satisfy the structure formation bounds on the fraction of non-cold dark matter \cite{Diamanti:2017xfo}. These bounds translated to the FIMP framework \cite{Kamada:2019kpe} read $\Omega_{\psi_1}/\Omega_\mathrm{DM} \lesssim 0.2$ at $3 \sigma$ confidence level  for $m_1 = 40\eV$, and get relaxed for larger $m_1$. For the values of parameters saturating the upper limit on the flux, $\Omega_{\psi_2}=2\Omega_{\psi_1}$, the warmness constraints are therefore marginally satisfied for $m_1=40$ eV, but well satisfied for larger masses.

\begin{figure}[bt]
	\begin{center}
		\includegraphics[width=0.55\textwidth]{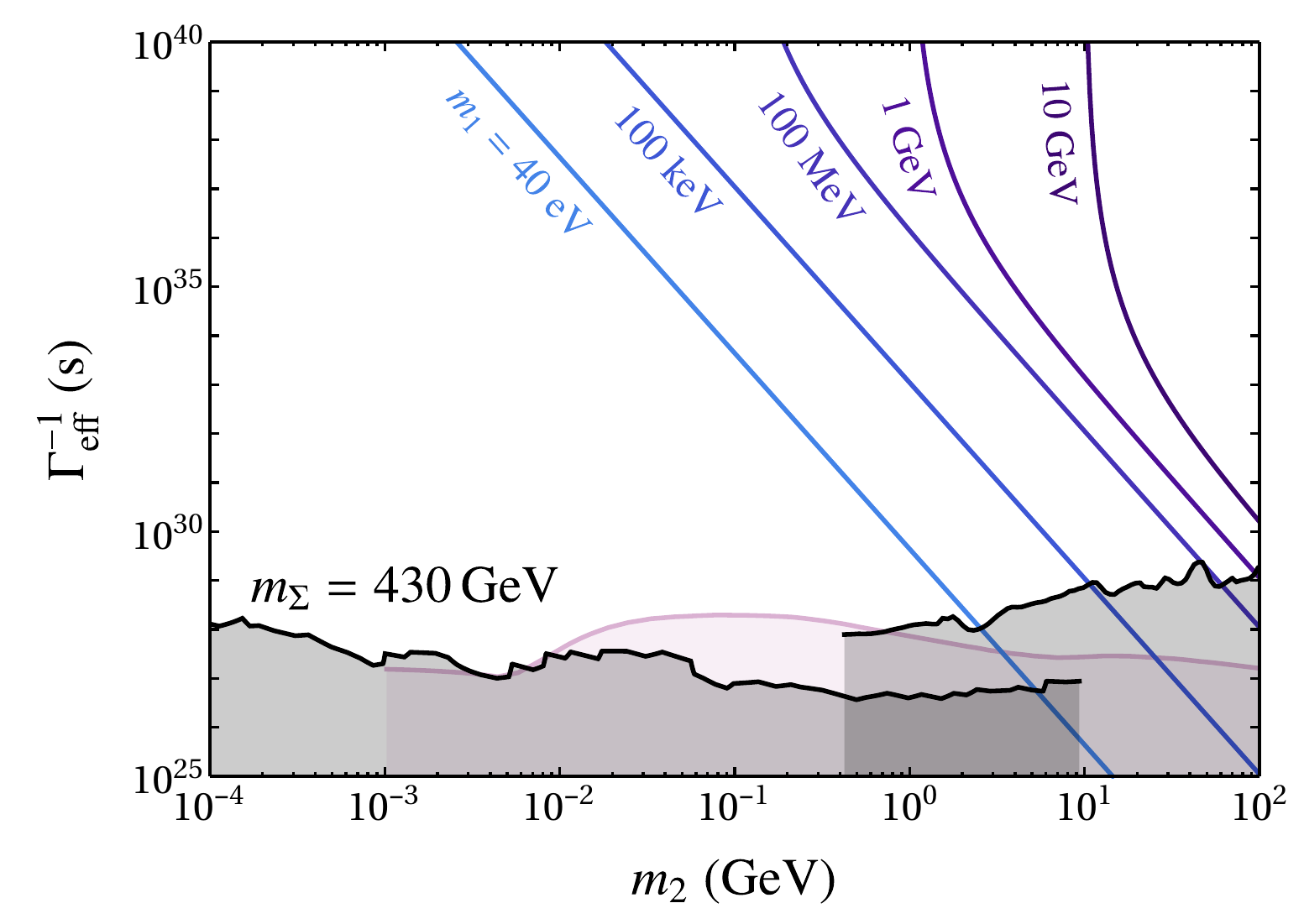}
	\end{center}
	\caption{\small Lower limit on the inverse decay rate for $\psi_2\rightarrow\psi_1\gamma$ in a multicomponent fermionic FIMP dark matter scenario, as a function of the mass of the decaying FIMP component $m_2$ for different values of the mass of the stable FIMP component $m_1$. The FIMP components are assumed to have a Yukawa coupling to a heavy scalar $\Sigma$ and to a right-handed electron. The grey regions correspond to the lower limit on the rate from the non-observation of a statistically significant sharp feature in the isotropic diffuse photon flux, and the pink regions to the recast limit on the rate from the non-observation of signatures of the decay $\psi_2\rightarrow \psi_1 e^+ e^-$ in CMB data. The mass of the mediator $\Sigma$ has been fixed to $430\GeV$.}
	\label{fig:fermionDecayResult}
\end{figure}

In Figure~\ref{fig:fermionDecayResult} we show the lower limit on the inverse width for the process $\psi_2\rightarrow\psi_1\gamma$ from the requirement that the FIMP density generated via freeze-in does not exceed the measured dark matter density, as a function of the mass of the decaying FIMP and for fixed values of the mass of the stable FIMP. For the latter, we require $m_1\gtrsim 40$ eV to ensure that $\psi_1$ is always out of thermal equilibrium.
In our analysis, we have fixed for concreteness $m_\Sigma=430$ GeV, which saturates the current lower limit on the mass of long-lived charged scalar particles, assuming Drell-Yan production \cite{Aaboud:2019trc}.
We show as dark gray shaded regions the values of the inverse width excluded by the non-observation of a line in the isotropic gamma-ray flux measured by INTEGRAL~\cite{Bouchet:2008rp}, COMPTEL~\cite{1999ApL&C..39..193W}, EGRET~\cite{Strong:2004de} and 
the Fermi-LAT~\cite{Ackermann:2014usa}, as calculated in~\cite{Essig:2013goa} (left) as well as a dedicated line-search by Fermi-LAT~\cite{Ackermann:2015lka} (right). Future gamma-ray telescopes like the proposed AMEGO~\cite{McEnery:2019tcm} can improve the sensitivity by an order of magnitude in the mass range $\sim 0.3-30$ MeV.
In the pink shaded area, gamma ray signals  are precluded in the present setup with $l=e$ by the constraints on the rate of the associated three body decay $\psi_2 \to \psi_1 e^+e^-$ derived in \cite{Slatyer:2016qyl} from the non-detection of an exotic energy injection in the thermal plasma during CMB decoupling, recast into limits on the decay rate into gamma rays using the results from \cite{Garny:2010eg}.

Notably, there are choices of parameters for which multicomponent fermionic FIMP frameworks could be probed by current experiments. This requires, {\it e.g.} $m_2\gtrsim 3$ GeV for $m_1=40$ eV,  $m_2\gtrsim 20$ GeV for $m_1=100$ MeV, or $m_2\gtrsim 90$ GeV for $m_1=1$ GeV. In all the cases, the spectrum of FIMP masses must be hierarchical to obtain appreciable gamma ray signatures.
Our limits are only conservative and are meant to illustrate that multicomponent FIMP dark matter scenarios may lead to observable signals in indirect searches; limits from the non-observation of features in the electron/positron spectrum or from the gamma-rays produced by inverse Compton in the propagation of the electrons/positrons in the interstellar medium could lead to competitive or even better constraints than the ones derived in this work. A comprehensive analysis of the signals from multicomponent FIMP scenarios is beyond the scope of this work.

\section{Multicomponent scalar FIMP from Higgs decay}
\label{sec:scalarFIMP-Higgs}
In this section we consider  a scenario where the Standard Model is extended with two scalar gauge singlets, $\phi_1$ and $\phi_2$, with masses $m_1$ and $m_2$, $m_2>m_1$, both odd under the same $\mathbb{Z}_2$ symmetry. With this set-up, $\phi_1$ is absolutely stable, and therefore a dark matter component, while $\phi_2$ decays into $\phi_1$, with a lifetime that depends on the model parameters.

In the minimal set-up, the interaction of $\phi_1$ and $\phi_2$ with the Standard Model particles occurs through the Higgs portal term $\lambda_{ij} H^\dagger H \phi_i\phi_j$, with $H$ the Higgs doublet. This term gives, upon electroweak symmetry breaking, the following cubic and quartic interactions:
\begin{align}
-{\cal L}_{\rm int}=\frac{1}{2}\lambda_{ij} v h \phi_i\phi_j+\frac{1}{2}\lambda_{ij} h^2 \phi_i\phi_j\;,
\end{align} 
with $h$ the Higgs boson and $v=\langle H^0\rangle/\sqrt{2}\simeq 174$ GeV. We assume $\lambda_{ij}$ very small, such that $\phi_1$ and $\phi_2$ are both FIMPs. 

In this scenario the heavier dark matter component can decay into the lightest through $\phi_2 \to \phi_1 \gamma \gamma$, producing a distinctive gamma-ray spectrum. The source term is given by Eq.~(\ref{eq:source-term}), in this case with a partial width given by 
\begin{equation}
\label{eq:scalarDecay2Photons}
\Gamma_{\phi_2 \to \phi_1 \gamma \gamma} = \frac{1}{26880 \pi^3} \left( \frac{\lambda_{12} c_{\gamma\gamma}}{m_h^2} \right)^2 m_2^5 \Delta^7 \,_2F_1\left(3,4,8;\Delta\right),
\end{equation}
and energy spectrum calculated in \cite{Ghosh:2019jzu}.
Here, $c_{\gamma\gamma} \simeq -2.03\times 10^{-3}$ is the effective coupling of the Higgs to two photons,  $\Delta = 1-m_1^2/m_2^2$ parametrizes the mass difference between both FIMPs,  and $\,_2F_1\left(3,4,8;\Delta\right)$ is a hypergeometric function which takes values between 1 and 35 for $\Delta$  between 0 and 1. The coupling $\lambda_{12}$ leads also to dark matter production via freeze-in, along with the couplings $\lambda_{11}$ and $\lambda_{22}$. Therefore, following the same rationale as in Section \ref{sec:fermionFIMP}, one expects an upper limit on the gamma-ray flux from the requirement that the FIMP density in our Universe does not exceed the measured dark matter density $\Omega_{\phi_1}+\Omega_{\phi_2}\leq \Omega_{\rm DM}$.

For $m_h>m_1+m_2$, FIMP production is dominated by the decay processes $h\rightarrow \phi_i\phi_j$, with decay rates given  by:
\begin{equation}
\Gamma_{h\to \phi_i \phi_j} = \frac{\kappa \lambda_{ij}^2 v^2}{16 \pi m_h} 
\sqrt{1 - \frac{4 m_2^2}{m_h^2} + \frac{2  m_2^2 \Delta}{m_h^2} + \frac{m_2^4 \Delta^2}{m_h^4}}\;,
\end{equation}
where $\kappa=1/2$ for $i=j$ and $\kappa=1$ otherwise.
Using Eq.~(\ref{eq:correctDecayRate}), and imposing the requirement $\Omega_{\phi_1}+\Omega_{\phi_2}\leq \Omega_{\rm DM}$, we obtain:
\begin{equation}
\label{eq:lambda12Decays}
\lambda_{12} \lesssim  1.2 \times 10^{-11} \left(\frac{m_2 (1+\sqrt{1-\Delta})}{\GeV}\right)^{-1/2} \left(1 - \frac{4 m_2^2}{m_h^2} + \frac{2  m_2^2 \Delta}{m_h^2} + \frac{m_2^4 \Delta^2 }{m_h^4}\right)^{-1/4}\;,
\end{equation}
with the upper limit being saturated if FIMP production is dominated by the channel $h\rightarrow \phi_1 \phi_2$.
The scalar FIMP masses $m_{1,2}$ are bounded by the same dark matter warmness constraints as in section~\ref{sec:fermionFIMP}.
Using Eq.~(\ref{eq:scalarDecay2Photons}) we obtain an upper limit on the decay width
\begin{equation}
\Gamma_{\phi_2 \to \phi_1 \gamma \gamma}\lesssim
\left(2 \times 10^{29} \,\mathrm{s} \right)^{-1}
\left(  \frac{m_2}{\MeV} \right)^4
\Delta^7\,_2F_1\left(3,4,8;\Delta\right) \;.
\label{eq:decay_rate_scalar}
\end{equation}
Finally, using the source term Eq.~(\ref{eq:source-term}) it follows that the flux of the diphoton signal is maximal when the decay rate saturates Eq.~(\ref{eq:decay_rate_scalar}) and the totality of the dark matter of the Universe was produced via $h\rightarrow\phi_1\phi_2$.

\begin{figure}[bt]
	\begin{center}
		\includegraphics[width=0.55\textwidth]{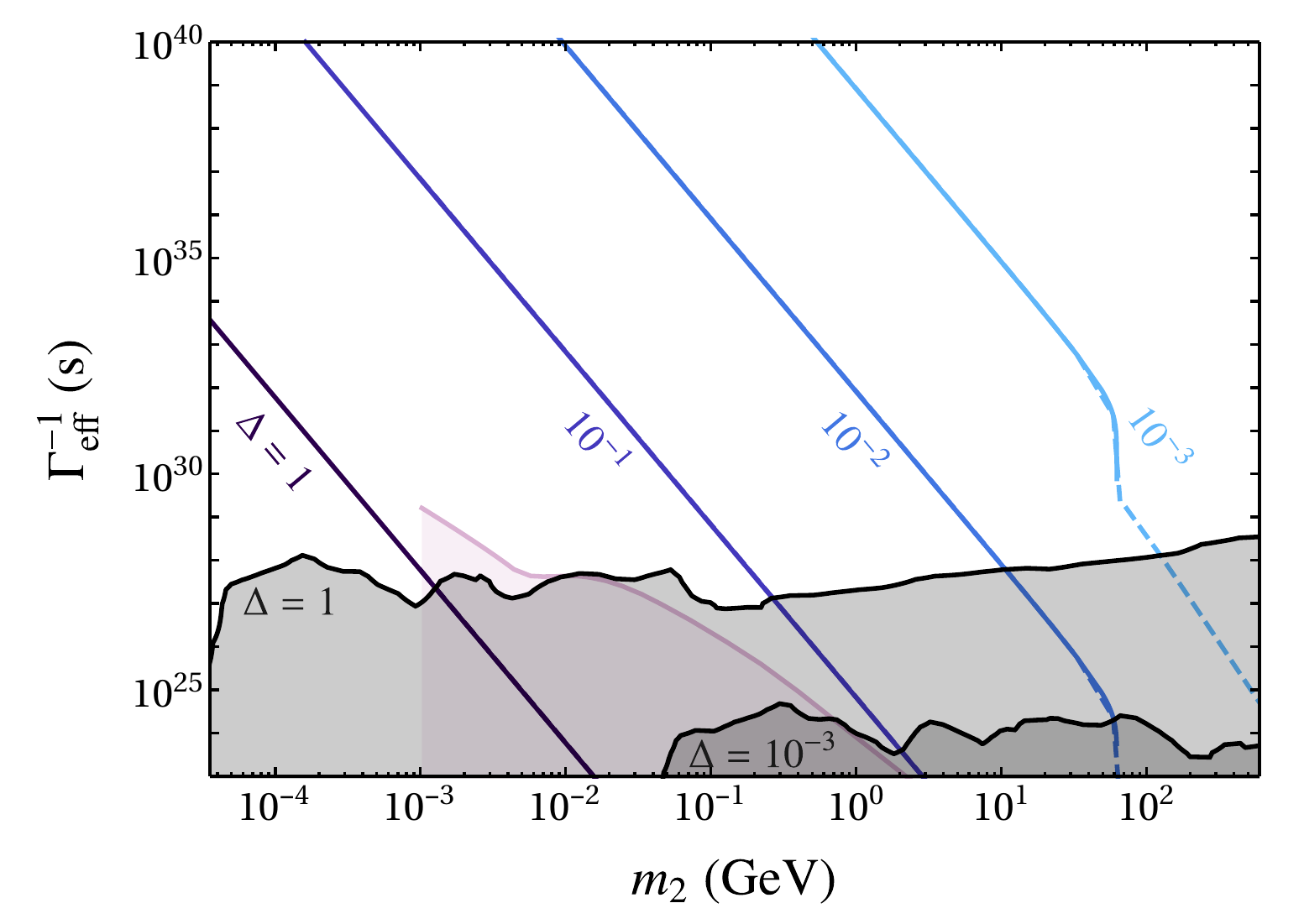}
	\end{center}
	\caption{\small Lower limit on the inverse decay rate for $\phi_2\rightarrow\phi_1\gamma\gamma$ in a multicomponent scalar FIMP dark matter scenario, as a function of the mass of the decaying FIMP component $m_2$ for different values of the degeneracy parameter $\Delta$. The FIMP components couple to the Higgs boson via a quartic coupling.  The grey regions correspond to the lower limit on the rate from the non-observation of a statistically significant sharp feature in the isotropic diffuse photon flux. In the pink region, gamma ray signals for $\Delta=1$ are precluded by CMB limits on the associated decay into electrons, as described in the text.
	}
	\label{fig:scalarDecayResult}
\end{figure}

We show in Fig.~\ref{fig:scalarDecayResult} our lower limit on the inverse width for the process $\phi_2\rightarrow \phi_1 \gamma\gamma$ as a function of the mass of the decaying dark matter component for different values of the degeneracy parameter $\Delta$. The dashed lines show the results obtained using \textsc{micrOMEGAs}, and the solid lines correspond to our analytical estimate Eq.~\ref{eq:decay_rate_scalar}. As expected from our previous discussion, the analytical estimate reproduces well the numerical upper limit when $m_1+m_2< m_h$.

We also show in the Figure the limits on the inverse width for $\phi_2\rightarrow\phi_1\gamma\gamma$ obtained in \cite{Ghosh:2019jzu}  from the non-observation of a statistically significant sharp feature in the isotropic diffuse photon flux~\cite{Bouchet:2008rp,1999ApL&C..39..193W,Strong:2004de,Ackermann:2014usa} 
%determined by INTEGRAL~\cite{Bouchet:2008rp}, COMPTEL~\cite{1999ApL&C..39..193W}, EGRET~\cite{Strong:2004de} and the Fermi-LAT~\cite{Ackermann:2014usa} 
(see also \cite{Essig:2013goa}); we show only the contours for $\Delta=1$ (very hierarchical spectrum) and for $\Delta=10^{-3}$ (very degenerate spectrum), as the limits are only mildly dependent on $\Delta$ in the degenerate case.
Signals are also expected for all other kinematically accessible decay channels of the off-shell Higgs, a detailed analysis of which we leave to future work. As an example, the pink shaded region is excluded by anomalous energy injection during CMB decoupling~\cite{Slatyer:2016qyl} through the process $\phi_2\rightarrow\phi_1e^+e^-$~\cite{Ghosh:2019jzu} only, illustrating the possibility of multiple complementary probes. The line shown is for the hierarchical $\Delta=1$ case; in the degenerate case, the limit lies outside of the figure.

It follows from the figure that this scenario may leave observable imprints in gamma rays when the mass of the decaying FIMP dark matter component larger than a few MeV.
In particular, the scenario where $|\lambda_{11}|,|\lambda_{22}|\ll |\lambda_{12}|$, such that FIMP production is dominated by the channel $h\rightarrow\phi_1\phi_2$, is excluded for hierarchical FIMP components when $m_2 \gtrsim 1 \MeV$. For degenerate FIMP components, the decay rate receives a phase-space suppression, thus avoiding the gamma-ray limits when $m_2\lesssim 2 \GeV$, $m_2\lesssim 55 \GeV$ or $m_2 \lesssim 800 \GeV$ when $\Delta=10^{-1}$, $\Delta=10^{-2}$ and $\Delta=10^{-3}$ respectively. The limits are also relaxed when $|\lambda_{12}|\ll |\lambda_{11}|,|\lambda_{22}|$, such that the decay process $\phi_2\rightarrow\phi_1\gamma\gamma$ is suppressed or $\phi_2$ is not the dominant dark matter component.

Let us note that the FIMP quartic coupling interactions with the Higgs boson $\lambda_{ij} H^\dagger  H \phi_i \phi_j$ leads not only to cubic and quartic portal interactions with the Standard Model, but also to a contribution to the FIMP masses. For freeze-in production, ${\rm max}(\lambda_{ii},\lambda_{ij})\sim 10^{-11} (m_i/{\rm GeV})^{-1/2}$. Therefore, the FIMP mass matrix receives a contribution from electroweak symmetry breaking which is $\delta m^2_{ij}\sim \lambda_{ij} v^2/2\sim 3.5$ MeV. Correspondingly, a scalar FIMP mass below this value typically requires special choices of parameters.

\section{Multicomponent scalar FIMP DM from heavy charged fermion decay}
\label{sec:scalarFIMP-loop}
We finally consider a variant of the previous scenario, where we assume the Higgs portal interactions to be negligibly small, and instead the two real scalar dark matter dark matter candidates $\phi_{1,2}$ couple to the Standard Model bath via a Yukawa coupling to a $\mathbb{Z}_2$-odd charged $SU(2)$-singlet fermion $\Psi$ and a standard model lepton. The relevant interaction term is 
\begin{equation}
\mathcal{L}_\Psi = y_{i} \phi_i \bar\Psi P_R l +\,\mathrm{h.c.}
\end{equation}

In this scenario the heavier dark matter component can decay into the lightest and two photons $\phi_2 \to \phi_1 \gamma \gamma$ through a loop involving the heavy fermion and the lepton. Expressions for the differential decay rate $d\Gamma_{\phi_2 \to \phi_1 \gamma \gamma}/dE_\gamma$ and the partial decay rate $\Gamma_{\phi_2 \to \phi_1 \gamma \gamma}$ are given in~\cite{Ghosh:2019jzu} and depend on the masses of both particles in the loop.

As in the previous sections, one finds that the decay rate of the process $\phi_2\rightarrow\phi_1\gamma\gamma$ is bounded from above by the requirement of not overproducing dark matter. 
The relic abundance of the FIMP DM components $\phi_1$ and $\phi_2$ generated in the decay  $\Psi\rightarrow \phi_i l$ can be calculated from eq.~(\ref{eq:freezeInAbundanceResult}), with a decay rate  given by:
\begin{align}
\label{eq:PsiDecayRate}
\Gamma_{\Psi\to\phi_i l} &= \frac{|y_{i}|^2 m_\Psi}{32 \pi} \left(1 + \frac{2 m_l}{m_\Psi} + \frac{m_l^2}{m_\Psi^2} - \frac{m_i^2}{m_\Psi^2}\right) \sqrt{1 - \frac{2(m_l^2 + m_i^2)}{m_\Psi^2} + \frac{(m_l^2 - m_i^2)^2}{m_\Psi^4}} \nonumber \\
&\simeq \frac{|y_{i}|^2 m_\Psi}{32 \pi} ,
\end{align}
where in the last line we have assumed $m_{\phi_i}, m_{l} \ll m_\Psi$. The requirement $\Omega_{\phi_1}+\Omega_{\phi_2}\leq \Omega_{\rm DM}$ translates into the limit on the couplings:
\begin{align}
|y_{1}|^2 \frac{m_1}{m_\Psi} + |y_{2}|^2 \frac{m_2}{m_\Psi} \leq 3.1 \times 10^{-24}\;.
\label{eq:scalar_relic_abundance}
\end{align}
The freeze-in assumption is satisfied for
\begin{equation}
|y_{i}| \ll 5 \times 10^{-9} \sqrt{\frac{m_\Psi}{\GeV}},
\end{equation}
with the same lower limit on the FIMP mass as in the fermion case, $m_\mathrm{FIMP} \gtrsim 40 \eV$. This scenario is also subject to structure formation constraints, as described in  section~\ref{sec:fermionFIMP}.

\begin{figure}[bt]
		\begin{center}
			\includegraphics[width=0.32\textwidth]{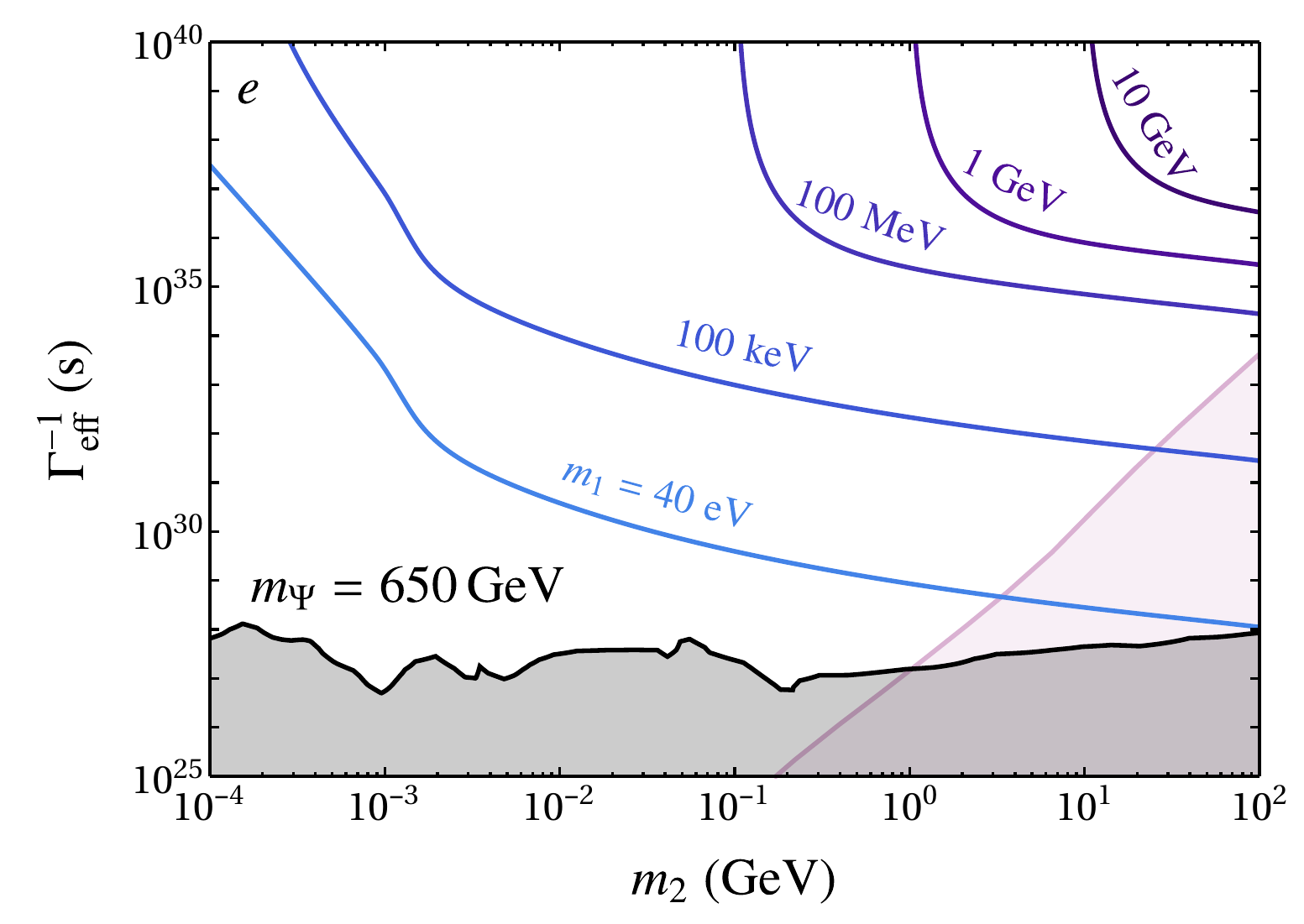}
			\includegraphics[width=0.32\textwidth]{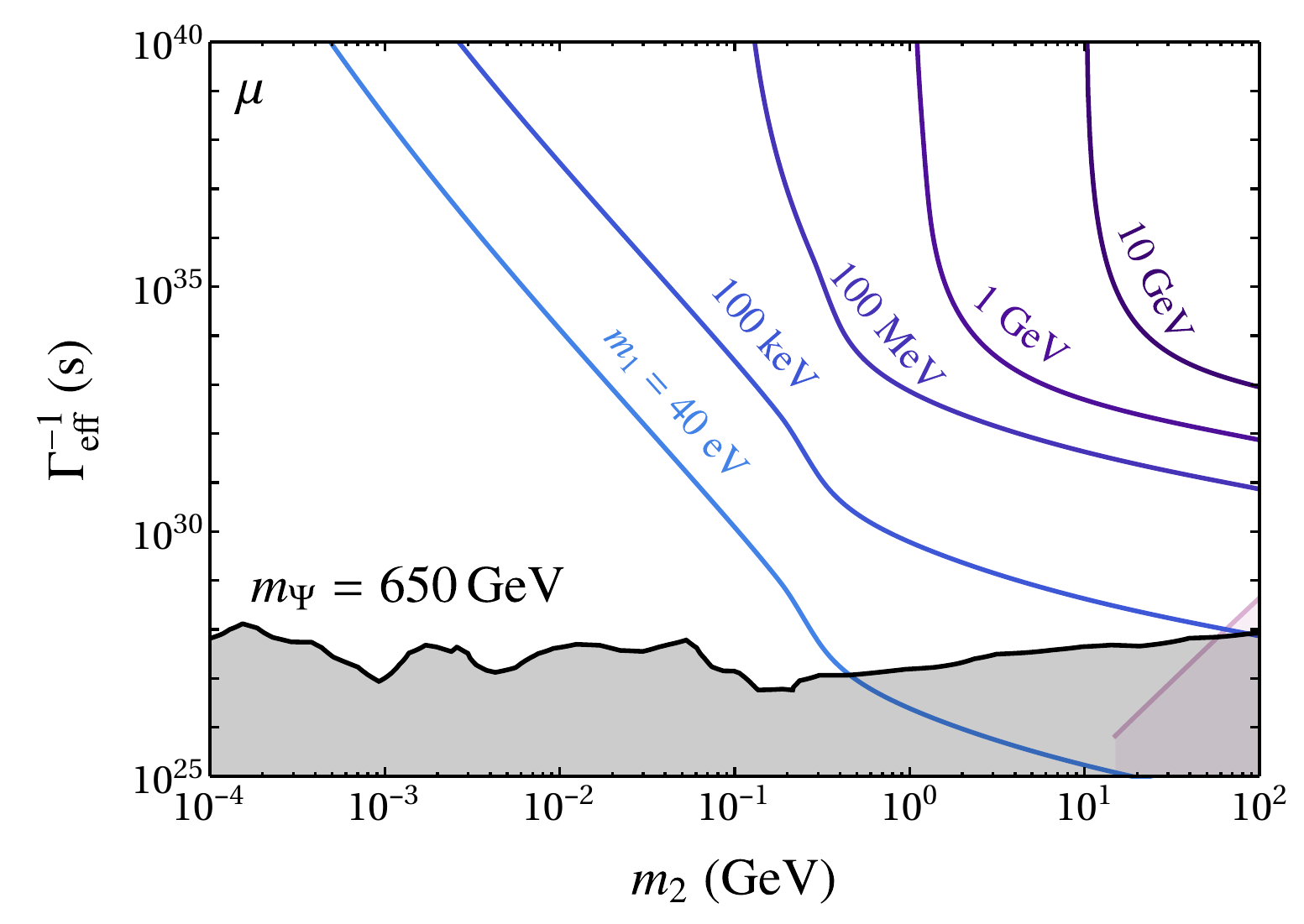}
			\includegraphics[width=0.32\textwidth]{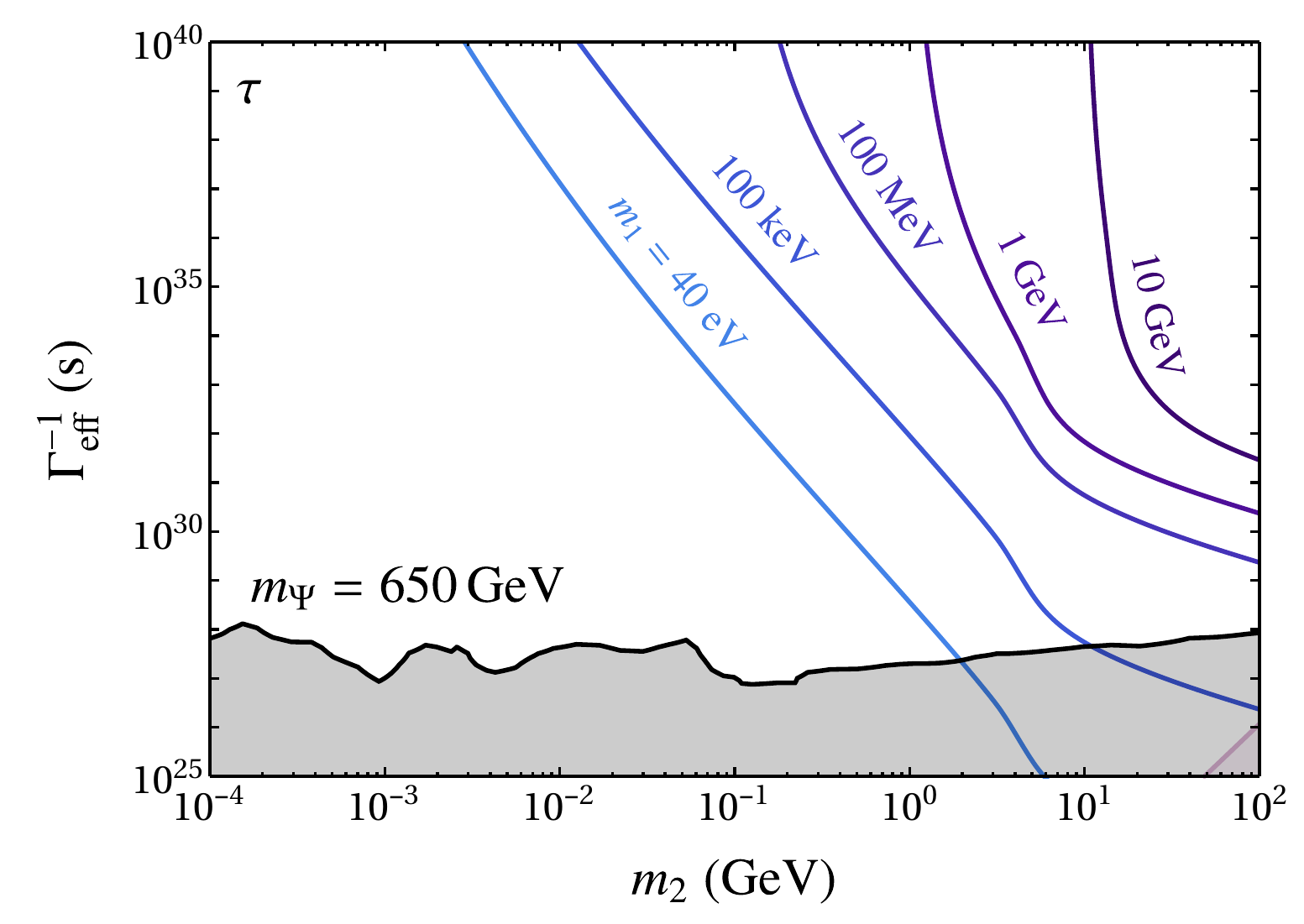}
		\end{center}
		\caption{\small Lower limit on the inverse decay rate for $\phi_2\rightarrow\phi_1\gamma\gamma$ in a multicomponent scalar FIMP dark matter scenario, as a function of the mass of the decaying FIMP component $m_2$ for different values of the mass of the stable FIMP component $m_1$. The FIMP components are assumed to have a Yukawa coupling to a heavy fermion $\Psi$ and to a right-handed electron (left panel), right-handed muon (middle panel) or a right-handed tau (right panel). Grey and pink regions are as in Fig.~\ref{fig:fermionDecayResult}. The mass of the mediator $\Psi$ has been fixed to $650\GeV$.}
		\label{fig:scalarWithMediatorDecayResult}
\end{figure}

In Figure~\ref{fig:scalarWithMediatorDecayResult} we show the lower limit on the inverse width for the process $\phi_2\rightarrow\phi_1\gamma\gamma$ from the requirement that the FIMP density generated via freeze-in does not exceed the measured dark matter density, as a function of the mass of the decaying FIMP and for fixed values of the mass of the stable FIMP.
In our analysis, we have fixed for concreteness $m_\Psi=650$ GeV, which saturates the current lower limit on the mass of long-lived charged fermions, assuming Drell-Yan production \cite{Khachatryan:2016sfv}. We show results for $l = e,\mu,\tau$, which due to their different masses give different predictions for the decay rate~\cite{Ghosh:2019jzu}.

We also show in the Figure the limits on the inverse width for $\phi_2\rightarrow\phi_1\gamma\gamma$ obtained in \cite{Ghosh:2019jzu} from the non-observation of a statistically significant sharp feature in the isotropic diffuse photon flux; we show only the contours for $\Delta=1$ (very hierarchical spectrum). The limits shown in Fig.~\ref{fig:scalarWithMediatorDecayResult} are analogous to those in Fig~\ref{fig:fermionDecayResult}, taking into account the slightly different gamma ray spectra compared to section~\ref{sec:scalarFIMP-Higgs} and the fact that appreciable decay rates are only possible for very hierarchical spectra.
As in Fig.~\ref{fig:fermionDecayResult}, we also show in pink limits on the diphoton decay rate from recasting limits from the non-observation of exotic energy injection from the tree-level decay $\phi_2 \to \phi_1 l^+ l^-$ during CMB decoupling corresponding to each scenario.

We find that gamma ray spectral features can be a sensitive probe to decaying leptophilic scalar FIMP dark matter if the dominant dark matter component $\phi_2$ is in the GeV range and the spectrum is very hierarchical. The lepton mass dependence of $\Gamma_{\phi_2 \to \phi_1 \gamma \gamma}$ results in more promising observational prospects if the dark sector couples to the $\mu$ or $\tau$ than to the electron. In the latter case, however, the observation of a gamma ray signal is precluded by the correlated tree-level decay $\phi_2 \to \phi_1 e^+e^-$, illustrating the potential of discerning among these scenarios by the complementarity of different search strategies.

\section{Conclusions}
\label{sec:conclusion}

Feebly Interacting Massive Particles (FIMPs) are well motivated dark matter candidates that can reproduce  the observed dark matter abundance through the slow decays of particles coupled to the Standard Model bath, a mechanism dubbed freeze-in. Unfortunately, in many models the minimal FIMP dark matter scenario is difficult to probe due to the tiny coupling of the FIMP to the Standard Model. On the other hand, the dark sector could be as rich and complex as our visible sector, and there might be more than one cosmologically long-lived FIMP in our Universe. In this work, we have argued that this scenario has new qualitative features compared to the minimal (single component) FIMP dark matter scenario, which may allow to probe the freeze-in mechanism.

To illustrate this idea, we have considered scenarios with two FIMP dark matter components such that the lighter component is absolutely stable, while the heavier component can decay into the lighter one and Standard Model particles.
We have shown that freeze-in production translates into an upper limit on the decay rate, thus setting a target for indirect dark matter searches, where signals may be found assuming freeze-in production.
This target is analogous to the well-known ``thermal'' annihilation cross-section $\langle\sigma v\rangle_{\rm th}$ in the weakly interacting massive particle (WIMP) paradigm, where experiments sensitive to $\langle\sigma v\rangle_{\rm th}\simeq 3\times 10^{-26}\,{\rm cm}^3/{\rm s}$ in a given annihilation channel are expected to find signals for particular model realisations.

Concretely, we have considered the decays $\psi_2\rightarrow\psi_1\gamma$ for fermionic FIMPs and $\phi_2\rightarrow\phi_1\gamma\gamma$ for scalar FIMPs (the decay $\phi_2\rightarrow\phi_1\gamma$ is forbidden by the conservation of angular momentum). These decays produce distinctive signals in the energy spectrum of the isotropic diffuse gamma-ray flux that can be easily distinguished from the featureless astrophysical background, leading to strong limits on these channels. We have calculated target decay rates for indirect searches in these scenarios, and we have found that some regions of their parameter space are already ruled out by current data.

Among the three scenarios analyzed in this work, the one with the best prospects of detection is the one with multicomponent scalar singlet dark matter, where both FIMPs couple to the Higgs boson via a quartic coupling.  In this scenario, FIMPs are produced via Higgs decays, and the heavy dark matter component decays into the lighter one through an off-shell Higgs boson. Assuming a hierarchical spectrum between the two dark matter components, current gamma-ray data allow to probe masses for the decaying component as low as 1 MeV. As the two dark matter components become more and more degenerate, only larger masses can be probed, in order to compensate the suppression of the decay rate by the smaller phase space. We have also analyzed scenarios where both FIMPs couple to a heavy mediator, which participates in freeze-in production and induces the decay of the heavier DM component. In these scenarios, the prospects for detection are less promising, yet current experiments can probe masses of the decaying dark matter component larger than $\sim$ 1 GeV. 

To summarize, we have demonstrated that some multicomponent FIMP dark matter scenarios can lead to detectable signals in the form of sharp features in the gamma ray energy spectrum.
Should such a signal be discovered in a future gamma-ray telescope, complementary searches for new particles, notably at the LHC, may allow to pin down the characteristics of this class of FIMP scenarios. This possibility will be investigated elsewhere.

\section*{Acknowledgments}
This work has been partially funded by the Collaborative Research Center SFB1258 and by the Deutsche Forschungsgemeinschaft (DFG, German Research Foundation) under Germany's Excellence Strategy – EXC-2094 – 390783311. JH wants to thank Laura Lopez-Honorez, Mathias Garny, Jan Heisig and Felix Kahlhoefer for helpful conversations.

\appendix

\normalem 
\bibliographystyle{JHEP}
\bibliography{refsPaper}
\end{document}